\documentclass[
reprint,
nofootinbib,
 amsmath,amssymb,
 aps,
pra,
twocolumn
]{revtex4-2}
\usepackage{graphicx}% Include figure files
\usepackage{dcolumn}% Align table columns on decimal point
\usepackage{bm}% bold math
\usepackage[export]{adjustbox}
\usepackage[english]{babel}

\usepackage{amsmath}
\usepackage{graphicx}
\usepackage[colorlinks=true, allcolors=blue]{hyperref}
\usepackage{braket}
\usepackage{mathtools}

\usepackage{natbib}
\usepackage{ulem}
\usepackage{xcolor}
\usepackage{xspace}

\newcommand{\HEi}{{\cal V}}
\newcommand{\amplitudeE}{A_E}
\newcommand{\sE}{\sigma_E}
\newcommand{\Np}{N_{\rm p}}
\newcommand{\Nc}{N_{\rm c}}
\newcommand{\rmi}{\mathrm{i}}
\newcommand{\rme}{{\rm e}}
\newcommand{\rmd}{{\rm d}}
\newcommand{\Up}{U_p}
\newcommand{\Fl}{F_L}
\newcommand{\wl}{\omega_L}
\newcommand{\Tl}{T_L}
\newcommand{\ti}{t_i}
\newcommand{\ts}{t_s}
\newcommand{\tr}{t_r}

\newcommand{\rhoe}{\hat{\rho}}
\newcommand{\rhoph}{\hat{\rho}_{\rm ph}}

\definecolor{jccol}{rgb}{0.2,0.8,0.2}
\newcommand{\LCPMR}{Sorbonne Universit\'{e}, CNRS, Laboratoire de Chimie Physique – Mati\`{e}re et Rayonnement, LCPMR, 75005 Paris, France}

\begin{document}
%\title{Strong-field dynamics in disordered environments : High Harmonic Generation in liquids \\ OU \\ Photoelectron dynamics in a disordered environment: High-order harmonic generation in liquids \\ OU \\ High-order harmonic generation from an atom in a disordered environment}
\title{High-order harmonic generation from an atom in a disordered environment}

\author{Simon His}
\author{Camille Lévêque}
\author{Jérémie Caillat}
\author{Richard Taïeb}
\author{Jonathan Dubois}
\email{jonathan.dubois@sorbonne-universite.fr}
\affiliation{\LCPMR}

%\author[1]{First Author}
%\author[2]{Second Author}
%\author[1,3,*]{Third Author}
%\affil[1]{\small Affiliation 1}
%\affil[2]{\small Affiliation 2}
%\affil[3]{\small Affiliation 3}
%\affil[ ]{\small *Corresponding author: email@example.com}

\begin{abstract}
\noindent
Using one-dimensional simulations analyzed through the lens of open quantum systems, we study the photoelectron's strong-field dynamics from an atom surrounded by a scattering environment stochastically structured. 
We theoretically investigate high-order harmonic generation from this situation.
We show that local dephasing of the photoelectron wavepacket induced by elastic scattering leads to global decoherence. This drives a transition from quantum to classical behavior, as witnessed by the photoelectron probability density localizing around specific trajectories of the classical analog system:  unstable periodic orbits. 
This phenomenon mirrors quantum scars traditionally observed in the eigenfunctions of time-independent systems, such as quantum billiards.
Here, it emerges in-situ within a time-dependent framework, manifesting directly in the real-time dynamics from the ground state rather than solely through spectral analysis.
\end{abstract}

\maketitle

\section{Introduction}

The discovery of high-order harmonic generation~\cite{Ferray1988} (HHG) from gaseous media has led to extensive fundamental and technological advances, such as the generation of attosecond light pulses~\cite{Paul2001, Hentschel2001} and the tracking in real time of electron dynamics in matter~\cite{Goulielmakis2010, Beaulieu2017, Autuori2022, Heldt2023}.
At the single-atom level, the most simple and widely accepted mechanism of this highly nonlinear phenomenon is the so-called three-step model~\cite{Krause1992, Corkum1993, lewenstein_theory_1994}, where a photoelectron wavepacket (i) undergoes tunnel ionization through the potential barrier created by the strong electric field, (ii) travels outside the atom guided by the oscillations of the driving field and returns to the parent ion, and (iii) recombines with the bound state, emitting high frequency radiations.
The goal of this study is to investigate how structural disorder, a fundamental characteristic of liquids, affects step (ii), using a simple model, which includes elastic scattering, designed to isolate this property from other contributions.

On the one hand, unlike gas, liquids have density and structural correlations that make it impossible to treat the problem as an isolated emitting atom. On the other hand, it is difficult to treat them with band structures, as one would for a crystalline solid, because of their inherent stochastic nature and associated loss of long-range structural order.
Despite these difficulties, time-dependent density functional theory (TDDFT) simulations have successfully reproduced several key differences between liquid- and gas-phase HHG~\cite{Neufeld2022, mondal_high-harmonic_2023, Mondal2025, Moore2025}. For example, TDDFT results suggest that the harmonic cutoff becomes independent of the driving field frequency beyond a specific intensity threshold--a phenomenon explained by a modified three-step model~\cite{mondal_high-harmonic_2023}.
Furthermore, these simulations have identified a secondary plateau, attributed to recombination in step (iii) that occurs at neighboring molecular sites rather than the ionized site~\cite{Mondal2025}.
These theoretical features are closely mirrored in experiments in which HHG spectra have been obtained from various liquids, such as water, methanol and ethanol, primarily using flat liquid jet setups~\cite{luu_extremeultraviolet_2018, Mondal2025}. However, while TDDFT reproduces the complex features observed in these experiments, it often acts as a `black box' where the individual contributions of elastic and inelastic scattering processes remain intertwined. This lack of mechanistic transparency suggests that neither brute-force simulations nor the elementary three-step model are sufficient, necessitating the development of intermediate reduced models to bridge the gap.

A notable attempt to bridge this gap involves simplified stochastic models based on one-dimensional dynamics, which were introduced to describe HHG in liquids~\cite{zeng_impact_2020}. It notably predicted a `transition harmonic'--a spectral boundary above which even harmonics emerge due to the breakdown of inversion symmetry. This was attributed to the short-range order of the liquid, with the transition frequency supposedly governed by the mean-free path and structural fluctuations. 
However, these specific spectral features have yet to be experimentally validated. This discrepancy likely stems from the model's reliance on an unconventional averaging over distinct initial eigenstates, an approach that may introduce numerical artifacts rather than reflecting the true physical ensemble. %Nevertheless, the existence of such a model highlights the potential for reduced-order frameworks to capture the influence of liquid-phase disorder, provided the underlying statistical ensemble is physically consistent.

We aim to establish a simple theoretical benchmark through one-dimensional simulations by revisiting the stochastic model in~\cite{zeng_impact_2020}.
We show that it allows us to reproduce a secondary plateau observed in the experiments~\cite{Mondal2025}.
In addition, we offer a novel approach by analyzing this complex process using an open quantum systems perspective. This approach naturally incorporates the structure of the environment through the use of statistically mixed states and reveals profound implications for the electron's behavior in step (ii). We find that the stochastic dephasing introduced in the electron's wave function by a scattering environment is a remarkable source of decoherence. 
This leads to a transition from quantum to classical chaotic motion, which has also been observed in systems in the presence of relaxations~\cite{zurek_decoherence_1991, habib_decoherence_1998}.
We show that this transition is accompanied by a localization of the probability density around periodic orbits of the classical analog system. This behavior is strikingly reminiscent of `quantum scars' typically observed in the spectral analysis of chaotic, time-independent systems~\cite{Heller1984, keski-rahkonen_quantum_2019}. 
However, our model occupies a unique middle ground: it is a closed quantum system, meaning it evolves unitarily without a traditional external reservoir, yet it is stochastic because the ensemble averaging over disordered spatial configurations. By bridging these two frameworks, we demonstrate that the underlying unifying theme remains the interplay between quantum and classical chaos, even in the presence of structural disorder.

The paper is organized as follows. In Section~\ref{sec:theory}, we describe the stochastic model employed to simulate an electron in an atomic potential subjected to both an intense laser field and a disordered environment. This section also details the numerical methods used to compute the various physical observables.
Section~\ref{sec:HHG_spectra} presents a comparative analysis of the HHG intensity spectra in the gas and liquid phases. Through the application of time-frequency analysis, we characterize the suppression of harmonic amplitudes and the emergence of scattering-induced spectral features.
In Section~\ref{analysis_photoelectron}, we establish a quantitative link between the harmonic peak intensities and the purity of the photoelectron state. We further demonstrate that the observed decoherence drives a dynamical transition from quantum to classical behavior, resulting in the localization of the wavefunction along unstable periodic orbits.
Finally, we summarize our findings and discuss their implications for ultrafast spectroscopy of disordered systems in Section~\ref{sec:conclusion}. Throughout this work, atomic units are used unless otherwise stated.

\section{Theoretical framework and methods \label{sec:theory}}

\subsection{Hamiltonian framework}

For clear physical interpretations, we aim to anchor our simulations in a well understood reference system that we then perturb with a model liquid environment. 
The total Hamiltonian is thus of the form
\begin{equation}
\label{eq:H_main}
    H (x,p,t ; \mathbf{X}) = H_0 (x,p,t) + \HEi (x ; \mathbf{X}) ,
\end{equation}
where $\HEi (x)$ describes the interaction with the environment and 
\begin{equation}
\label{eq:H0}
    H_0 (x,p,t) = \frac{p^2}{2} + V(x) + x F(t)  ,
\end{equation}
corresponds to the usual one-dimensional Hamiltonian. It was already used for the description of the HHG response of a single one-dimensional atom~\cite{lewenstein_theory_1994, Peng2015}. The polarized electric field is
\begin{equation}
\label{eq:Ft}
    F(t) = \Fl f(t) \sin ( \wl t) ,
\end{equation}
with amplitude $\Fl$, frequency $\wl$ and envelope $f(t)$. 
In this article, we consider a laser intensity $7.896 \times 10^{14} \; \rm W \ cm^{-2}$ (0.25 a.u.), a laser wavelength of 1030 nm ($\wl = 0.044$ a.u.) and a trapezoidal envelope $f(t)$ with 2 laser cycle ($\Tl = 2\pi/\wl$) ramp-up, 11 laser cycle plateau and 2 laser cycle ramp down.
The electron-ion interaction is described by the soft-Coulomb potential~\cite{Javanainen1988} $V (x) =  - ( x^2 + 0.4837 )^{-1/2}$ leading to an ionization potential $I_p = 0.90$ a.u., close to that of He.

To model the interaction with the environment, we consider $\Np$ perturbers positioned at $x_k$ and such that $\mathbf{X} = \lbrace x_1 \dots x_{\Np} \rbrace$, on which the photoelectron can elastically scatter.
It is formally given by 
\begin{equation}
\label{eq:HE}
    \HEi (x ; \mathbf{X}) = \sum_{k=1}^{\Np} v ( x - x_{k})  ,
\end{equation} 
where each perturber, located at $x_k$, is modeled by the same \textit{attractive} Gaussian well
\begin{equation}
    v (x) = - \amplitudeE \, \exp \left( - \frac{x^2}{2\sE^2} \right) ,
\end{equation}
which represents an \textit{electro-negative} element.
This description in terms of perturbers is also motivated by theoretical studies on Rydberg molecules~\cite{Eiles2016, Hunter2020} or quantum scars~\cite{keski-rahkonen_quantum_2019}. 
For a given configuration of the environment, the electron wavefunction is a pure state $\ket{\psi (t;\mathbf{X})}$ conditioned by $\mathbf{X}$.
Its time-evolution follows the time-dependent Schr\"{o}dinger equation (TDSE)
\begin{equation}
\label{eq:TDSE}
    {\rm i} \partial_t \ket{\psi (t ; \mathbf{X})} = \hat{H} (t; \mathbf{X}) \ket{\psi (t ;  \mathbf{X})} ,
\end{equation}
with $\hat{H} (t;\mathbf{X})$ the Hamilton operator associated with~\eqref{eq:H_main}.
It is solved numerically with a high-order symplectic integrator called BM4~\cite{bandrauk_complex_2006, MacLachlan2022}. 
The gas phase is obtained by setting $\amplitudeE = 0$ in Eq.~\eqref{eq:HE}, and we use $\Nc = 10^3$, $\amplitudeE = 0.8$ a.u. and $\sE = 0.5$ a.u. for the liquid phase.

\subsection{Ensemble modeling of structural disorder}

\begin{figure}
    \centering
    \includegraphics[width=.45\textwidth]{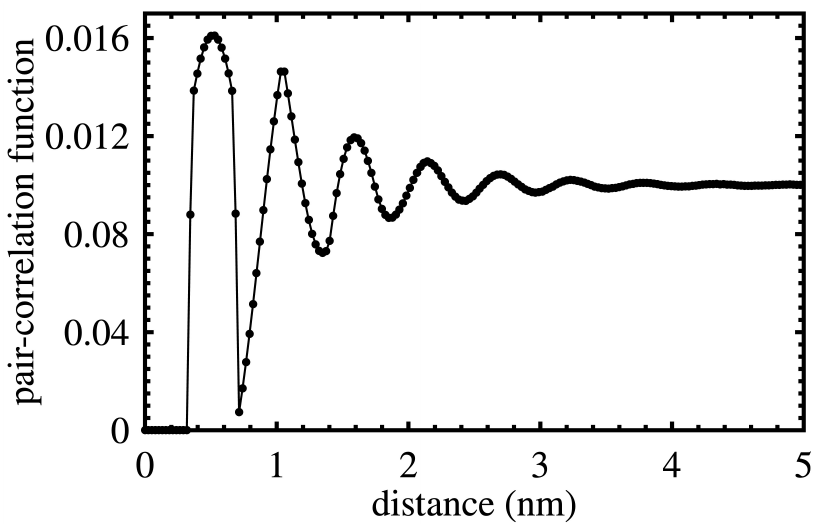}
    \caption{Pair-correlation function, following the joint probability distribution~\eqref{eq:fdx}, as a function of the distance between two scatterers.}
    \label{fig:gt}
\end{figure}

This system allows us to call upon a density matrix formalism for the description of the photoelectron state in the disordered environment. 
It can be easily understood as mixed states, each resulting from a given environment configuration such that
\begin{equation}
\label{eq:rho_general}
    \rhoe (t) = \int \ket{\psi (t ; \mathbf{X})} W(\mathbf{X})  \bra{ \psi (t ; \mathbf{X})} \; \rmd \mathbf{X} ,
\end{equation}
with $W(\mathbf{X})$ the joint probability distribution of the scatterer positions. This reflects the fact that each parent atom and electron sees a different and random environment.
The expectation value of an observable described by the operator $\hat{\cal O}$ is then obtained from
\begin{equation}
\label{eq:tr_rho}
    \braket{\hat{\cal O}} \equiv {\rm tr} \left[ \rhoe (t) \hat{\cal O} \right].
\end{equation}
For the scatterer configuration, we assume that the distance between two successive perturbers, $\Delta x_k = x_{k+1}-x_k$, follows a truncated Gaussian distribution~\cite{zeng_impact_2020}
\begin{equation}
    W_0 (x) = 
    \begin{dcases}
      \dfrac{1}{\cal N} \exp \left[ -\frac{\left( x - a \right)^2}{2\sigma^2} \right] , & x = \lbrack 2a/3, 4 a/3 \rbrack , \\
      0 ,           & \text{otherwise} ,
    \end{dcases}
\end{equation}
where ${\cal N}$ is the normalization constant. This spatial constraint prevents the overlap of the perturbers.
We consider an even number of perturbers $\Np$, distributed symmetrically around the origin. 
Let $j = \Np/2$ denote the index of the last perturber on the left.
The total joint probability distribution for the configuration $\mathbf{X}$ is given by
\begin{equation}
\label{eq:fdx}
    W(\mathbf{X}) =  W_0 (- x_j) W_0 (x_{j+1}) \prod_{k \in {\cal K}} W_0 (\Delta x_{k}) ,
\end{equation}
where ${\cal K} = \lbrace 1 , \ldots , \Np - 1 \rbrace \backslash \lbrace j \rbrace$.
The first two terms define the positions of the nearest neighbors of the parent ion. By constraining $- x_j$ and $x_{j+1}$ to the range $\lbrack 2a/3,4 a/3 \rbrack$, we ensure a `buffer zone' around the origin that prevents the bound state from being significantly distorted by the environment.
Throughout the article, we use a mean inter-scatterer distance of $a=10$ a.u. and a standard deviation of $\sigma = 1$ a.u. (0.529 and 0.0529 nm, respectively) unless otherwise stated. 

Figure~\ref{fig:gt} presents the pair correlation function derived from~\eqref{eq:fdx}, representing the distribution of distances $| x_k - x_j |$ for all pairs of perturbers. At short range, the distribution exhibits distinct peaks reminiscent of solvation shells, while the long-range behavior becomes uniform, characteristic of the structural disorder found in liquids~\cite{Ryu2020}. Note that the truncation of $W_0 (x)$ introduces nonphysical cusps in the distribution and does not alter our analysis that focuses primarily on the dichotomy between short- and long-range order. For the laser parameters employed here, the quiver radius ($\Fl / \wl^2$) is approximately 4 nm ($\approx 77$ a.u.), ensuring that the photoelectron extensively explores the environment across both local and macroscopic scales.

In practice $N_{\rm c}$ configurations $\mathbf{X}_i$ are generated randomly according to $W$ in Eq.~\eqref{eq:fdx}. The density matrix~\eqref{eq:rho_general} is thus approximated by
\begin{equation}
\label{eq:rhot}
    \rhoe (t) \approx \frac{1}{\Nc} \sum_{i=1}^{\Nc} \ket{\psi (t ; \mathbf{X}_i)}  \bra{ \psi (t; \mathbf{X}_i)} .
\end{equation}
As a result, the expectation value of $\hat{\cal O}$ is an incoherent sum over its expectation value for a given configuration, and Eq.~\eqref{eq:tr_rho} becomes
\begin{equation}
\label{eq:meanA}
    \braket{\hat{\cal O}} \approx \dfrac{1}{\Nc} \sum_{i=1}^{\Nc} \bra{\psi (t; \mathbf{X}_i)} \hat{\cal O} \ket{\psi (t; \mathbf{X}_i)} .
\end{equation}
In the following, we compare different observables for the gas and liquid phases.

\section{Influence of disorder on HHG spectra \label{sec:HHG_spectra}}

\subsection{HHG intensity spectrum}

\begin{figure}
    \centering
    \includegraphics[width=.45\textwidth]{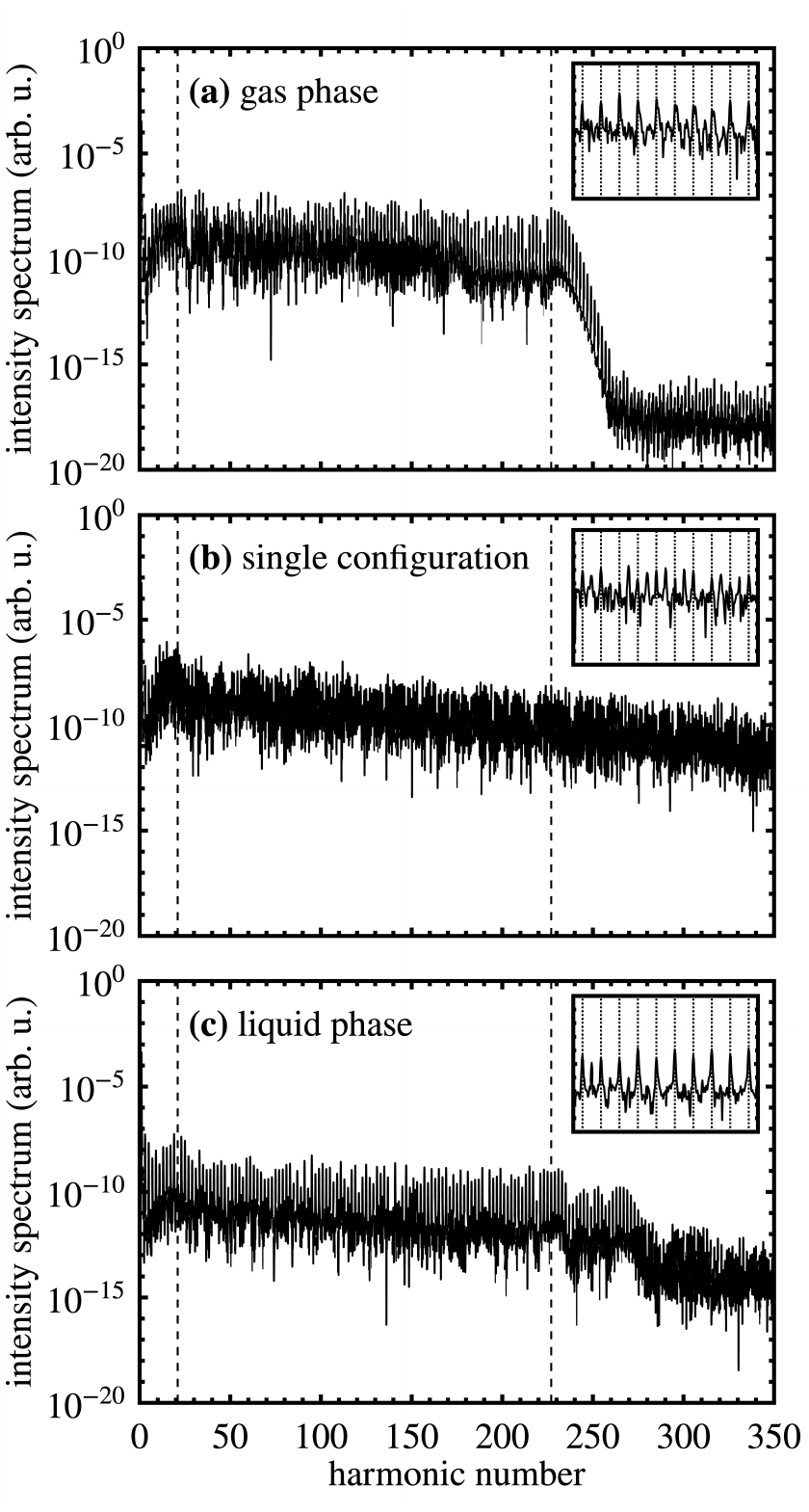}
    \caption{HHG intensity spectra obtained from the modulus of the Fourier transform of~\eqref{eq:dipole} as a function of the harmonic number for a laser intensity $7.896 \times 10^{14} \; \rm W \ cm^{-2}$, a laser wavelength of $1030$ nm, and a 2-11-2 trapezoidal envelope. (a) Gas phase with $\amplitudeE =0$ in Eq.~\eqref{eq:HE}. (b) Single configuration of the disordered environment ($\Nc = 1$) and (c) liquid phase with a thousand configurations ($\Nc = 10^3$), with $\amplitudeE = 0.8$ a.u. and $\sE = 0.5$ a.u. 
    The vertical dashed lines indicate harmonics 21 and 227. The insets are zooms on the harmonics $[150,170]$ and intensities $[10^{-15},10^{-5}]$. The vertical dotted lines indicate the odd harmonics.}
    \label{fig:HHG_spectrum}
\end{figure}

We start by analyzing the HHG intensity spectrum obtained from the dipole acceleration~\cite{Baggesen2011}. The latter is computed as
\begin{equation}
\label{eq:dipole}
    d (t) = \partial_t^2 \braket{\hat{x}} ,
\end{equation}
with $\braket{\hat{x}}$ evaluated from Eq.~\eqref{eq:meanA} and leading to an analytic form in terms of the derivatives of the potential according to Ehrenfest's theorem.
Figure~\ref{fig:HHG_spectrum} shows the HHG spectrum, obtained as the modulus of the Fourier transform of $d (t)$. We compare the results for the gaz phase (panel a), for a single configuration of the disordered environment (panel b), and for the liquid phase (panel c).
Figure~\ref{fig:HHG_spectrum}a shows that our model recovers the characteristic features of the gas-phase HHG spectra. The spectrum is dominated by odd-order harmonics and exhibits three distinct regions: a low-energy regime below the ionization potential $I_p$, a plateau extending to the semi-classical cutoff~\cite{Krause1992, Corkum1993, lewenstein_theory_1994} at $3.17 \ \Up + I_p$ around the 227th harmonic [$\Up = (\Fl / 2 \wl )^2$ the ponderomotive energy] and a sharp intensity roll-off beyond this limit.

% odd-even harmonics
The single-configuration spectrum in Fig.~\ref{fig:HHG_spectrum}b is notably structureless and contains both odd and even harmonics, reflecting the local loss of inversion symmetry induced by the disordered environment. In contrast, the liquid-phase spectrum (Fig.~\ref{fig:HHG_spectrum}c) exhibits exclusively odd harmonics. This restoration of symmetry arises because the ensemble average over all realizations recovers an effective macroscopic centrosymmetry. This mechanism is reminiscent of HHG in hetero-atomic molecular gases, where rotational averaging of randomly oriented molecules results in a purely odd-harmonic spectrum~\cite{Marchetta2025}.

% decrease of the overall intensity spectrum
For a single configuration, the spectrum decreases monotonically, losing nearly one order of magnitude between the 21st ($2\times 10^{-7}$) and 227th harmonics ($4\times 10^{-9}$). At low orders, the HHG intensity is similar to that of the gas-phase baseline, indicating that the ionization rate is largely unaffected by the environment. In contrast, the liquid phase exhibits a more uniform plateau, although with a globally suppressed intensity of roughly two orders of magnitude, as clearly illustrated in the insets of Fig.~\ref{fig:HHG_spectrum}.

% cutoff 
The liquid-phase spectrum exhibits a primary cutoff near the 227th harmonic, which, in our model, remains consistent with the gas-phase scaling ($3.17 \ \Up + I_p$). This observation invites a nuanced comparison with the results of Ref.~\cite{mondal_high-harmonic_2023}, where experimental measurements and TDDFT simulations indicated a primary cutoff that becomes independent of the driving frequency above a specific intensity threshold. In that work, the environment-induced mean free path was suggested to limit the trajectory length, preventing the electron from reaching the high kinetic energies predicted by the standard $\Up$ scaling.
Interestingly, despite our 1D geometry, which inherently maximizes the scattering probability by preventing the electron from bypassing perturbers, we do not recover this frequency-independent saturation. This suggests that the interplay between the mean free path and the recollision physics may be more complex than a simple truncation of trajectories. Although scattering events are frequent in our model, they do not necessarily suppress the high-energy return of the electron to the primary ion.

Furthermore, we observe a secondary plateau extending to the 267th harmonic, a feature reminiscent of the multi-plateau structures discussed in Ref.~\cite{Mondal2025} and attributed to off-site recombination at neighboring sites. The fact that our reduced model reproduces this extension indicates that environmental scattering can play a dual role: while it may influence the efficiency or scaling of the primary cutoff as reported in~\cite{mondal_high-harmonic_2023}, it simultaneously enables higher-energy recombination channels that extend the spectral range beyond the gas-phase limit. This persistence of multi-plateau features, even in a simplified 1D framework, underscores the robustness of scattering-assisted trajectories as a hallmark of liquid-phase HHG. To further resolve the timing and origin of these primary and secondary emission events, we turn to a time-frequency analysis in the following section.

\subsection{Time-frequency analysis}

\begin{figure}
    \centering
    \includegraphics[width=.45\textwidth]{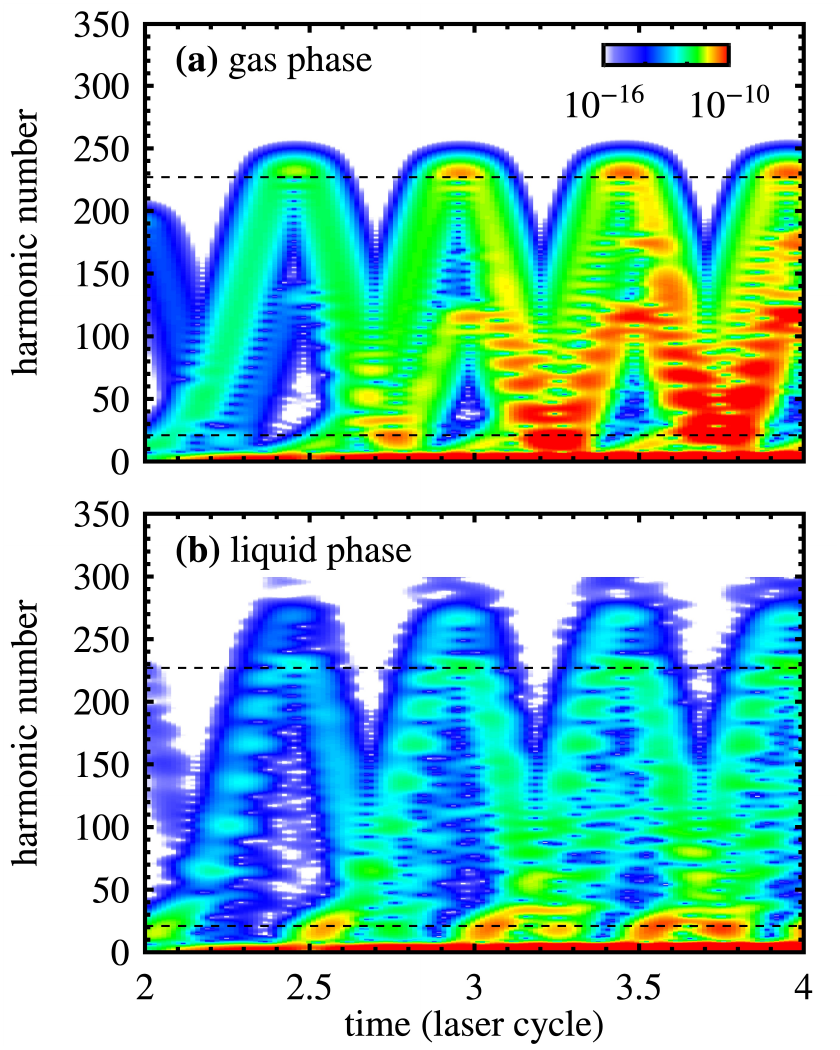}
    \caption{Time frequency analyses of the harmonic of the dipole [see Eq.~\eqref{eq:Gabor}] for (a) the gas phase and (b) the liquid phase. The parameters are the same as in Fig.~\ref{fig:HHG_spectrum}.}
    \label{fig:Gabor}
\end{figure}

Figure~\ref{fig:Gabor} shows a time-frequency map of the dipole acceleration $d(t)$ in Eq.~\eqref{eq:dipole} for the gas phase (panel a) and the liquid phase (panel b). 
It corresponds to the modulus of the Gabor transform
\begin{equation}
\label{eq:Gabor}
    \tilde{d}(\tau, \omega) = \int_{-\infty}^{\infty} d (t) \: w (\tau - t)  \: \rme^{- \rmi \omega t} \; \rmd t ,
\end{equation}
with a window function~\cite{Tate2007, berman_coherent_2018}
\begin{equation}
    w(t) = 
    \begin{dcases}
      \cos^4\left( \frac{\pi t}{T_{w}}\right) , & |t| < T_w/2 , \\
      0 ,           & \text{otherwise} .
    \end{dcases}
\end{equation}
We have used a window duration of $T_w = 0.35 \, \Tl$. 
The time-frequency analysis reveals that while the fundamental structure of the gas-phase emission--characterized by the classic short and long trajectories--is largely preserved in the liquid phase (Fig.~\ref{fig:Gabor}a-b), the introduction of disordered scattering neighbors induces significant changes. 
To provide a clearer framework for comparing the observed similarities and differences, we adopt a semi-classical approach based on the strong-field approximation (SFA). Using the simpleman model~\cite{Corkum1993, Tate2007, Koval2007, Mondal2025}, we can intuitively trace the photoelectron trajectories and their resulting emission. 

\begin{figure}
    \centering
    \includegraphics[width=.45\textwidth]{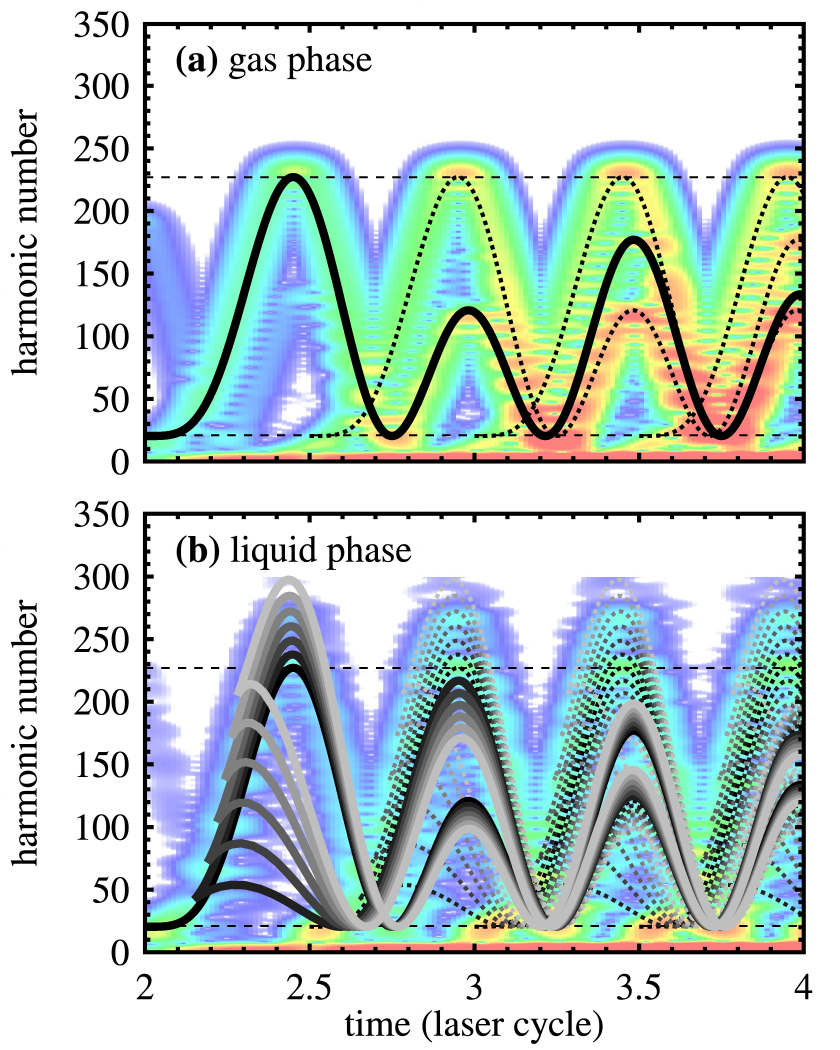}
    \caption{Same as Fig.~\ref{fig:Gabor} but with the kinetic energies of the electron as a function of the recombination time in the SFA. The solid lines are for ionization during the first laser cycle after the plateau. The dotted line are for later times, i.e. they correspond to the solid line translated by integer multiples of the half-period. The different colors correspond to different distances of return site with $|x(\tr)| = \ell$: black is for the return at $\ell = 0$, dark gray is at $\ell = a$ the mean inter-scatterer distance, lighter gray is $\ell = 2 a$, etc.}
    \label{fig:Gabor_trajectories}
\end{figure}

We thus assume that the electron is driven exclusively by the laser field [$V = 0$ in~\eqref{eq:H0}] upon ionization at time $\ti$, starting from the origin with zero initial kinetic energy, represented by $x(\ti) = 0$ and $p (\ti) = 0$.
In a constant-envelope field, the integration of Hamilton’s equations yields the trajectory
\begin{equation}
\label{eq:xt}
    x(t) = - \dfrac{\Fl}{\wl} \cos (\omega \ti) (t -\ti) + \dfrac{\Fl}{\wl^2} \left[ \sin (\omega t) - \sin (\omega \ti) \right]  .
\end{equation}
We analyze three different recollision pathways and compare them with the time-frequency maps in Fig.~\ref{fig:Gabor}.

We first consider recombination at a distance $\ell$ from the parent ion, where $\ell=0$ corresponds to the standard gas-phase return~\cite{Corkum1993} and $\ell \neq 0$ represents `off-site' recombination at a perturber~\cite{Mondal2025}. The return kinetic energy is given by
\begin{equation}
    E_r (\tr , \ti) = \dfrac{p(\tr)^2}{2} = 2 \Up \left[ \cos (\omega \tr) - \cos (\omega \ti) \right]^2 ,
\end{equation}
where $\tr$ is the time at which $|x(\tr)| = \ell$.
In the gas phase (Fig.~\ref{fig:Gabor_trajectories}a), $x(\tr) = 0$ leads to the well-known primary returns and higher-order secondary pathways~\cite{Tate2007, Koval2007}. In the liquid phase (Fig.~\ref{fig:Gabor}b), these higher-order returns are largely suppressed. By plotting the return conditions for $\ell = k a$ (with $k=1,\ldots,6$), we observe that off-site trajectories occur at similar times and energies as gas-phase returns. The incoherent summation of these multiple pathways leads to rapid decoherence and a blurred spectral structure.
Crucially, each distance $\ell$ defines a new maximum return energy. For distances small compared to the quiver radius, we find numerically that this maximum scales linearly
\begin{equation}
    E_{\max} (\ell) \approx 3.17 \, \Up  + 0.31 \, \Fl \, \ell .
\end{equation}
This relation explains the extension of the HHG cutoff observed in the liquid phase. Physically, the perturber acts as a site for a `pseudo-bound state'. Similarly to scattering HHG experiments, where trajectories are trapped near a core due to phase-space `stickiness'~\cite{Sand1999, berman_coherent_2018, Berman2019, Mauger2012}, the interference between the high-energy wavepacket and the one trapped around the perturber site triggers the emission observed above the standard $3.17 \ \Up$ limit.

\begin{figure}
    \centering
    \includegraphics[width=.45\textwidth]{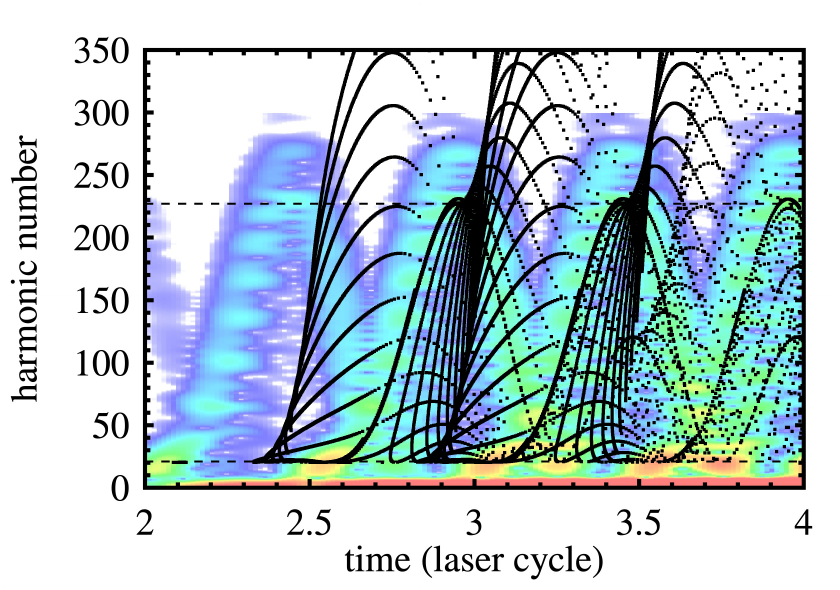}
    \caption{Same as Fig.~\ref{fig:Gabor_trajectories} but with return kinetic energies of backscattered photoelectrons as a function of the return time. The results are shown for electrons ionized in the first laser cycle after the end of the ramp-up of the laser envelope. The colormap is for the liquid phase in Figs.~\ref{fig:Gabor}b and~\ref{fig:Gabor_trajectories}b.}
    \label{fig:Gabor_BS_trajectories}
\end{figure}

A third channel involves electrons that backscatter off a perturber at time $\ts$ before returning to the origin. This is modeled by a momentum reversal $p (\ts) \to -p (\ts)$~\cite{Paulus1994}. This event acts as a `momentum reset': by reversing the direction at a strategic phase, the field can continue to accelerate the electron rather than decelerating it.
Although a single backscattering event can theoretically boost the return energy well beyond $3.17 \ \Up$, our time-frequency maps show that these conditions do not consistently match the dominant emission patterns (Fig.~\ref{fig:Gabor_BS_trajectories}). Consequently, while backscattering contributes to overall decoherence and the formation of a decaying spectral tail, its contribution to the structured HHG signal is negligible compared to off-site recombination, consistent with recent findings~\cite{Mondal2025}.

The dense and disordered environment of the liquid phase fundamentally reshapes the emission. The overlap of various off-site channels drives a general decrease in amplitude and `blurs' the trajectories in the time-frequency domain. However, the emergence of the secondary plateau remains a robust hallmark of the environment, confirming that the liquid phase enables trajectories that extend the spectral range beyond the predictions of the standard three-step model.
In the gas phase, the coherence of the emission is maintained by the uniqueness of the recollision path. In contrast, the liquid phase introduces a manifold of competing spatial channels, each with slightly different phases and timings, which interfere destructively when averaged. Next, we further analyze the photoelectron dynamics to quantify the effect of this decoherence through the purity of the state.

\section{Analysis of the photoelectron dynamics \label{analysis_photoelectron}}

\subsection{Purity of the photoelectron's state}

\begin{figure}
    \centering
    \includegraphics[width=.45\textwidth]{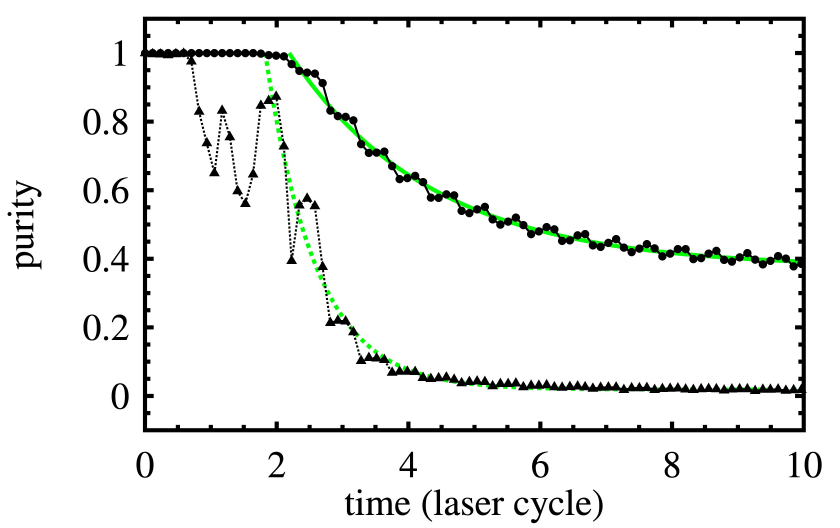}
    \caption{Purities of the total density matrix $P [ \rhoe (t)]$ (solid line with circles) and the free electronic part $P [ \rhoph (t)]$ (dotted line with triangles) computed with Eq.~\eqref{eq:purity} as a function of time. The green curves are the fit function from Eq.~\eqref{eq:purity_fit}. The parameters are the same as in Fig~\ref{fig:HHG_spectrum}.}
    \label{fig:purity_time}
\end{figure}

We now focus on the direct impact of these interactions on the quantum state of the electron and its transition to a classical statistical mixture. 
Although the total electronic state $\rhoe (t)$ includes both bound and free components, we isolate the contribution of the photoelectron $\rhoph (t)$ by applying a mask function. This mask suppresses the density matrix within 5 a.u. of the origin, effectively removing the ground-state contribution.
To account for both this masking and the loss of electron density that escapes the numerical grid, 
the degree of decoherence within the electronic system can be quantified by computing the normalized purity~\cite{Breuer2007}
\begin{equation}
\label{eq:purity}
    P [\rho (t)] = \frac{{\rm tr} [ \rho (t)^2]}{{\rm tr} [\rho (t)]^2} ,
\end{equation}
where $P [\rhoe (t)]$ and $P [\rhoph (t)]$ represent the purity of the electron and photoelectron, respectively.
A purity of 1 represents a pure quantum state, where the system remains fully coherent. A decrease in $P [ \rho (t) ]$ tracks the evolution of the photoelectron to a mixed state, signaling that the environment interactions have transformed the state into a distribution governed by classical statistics.

Figure~\ref{fig:purity_time} illustrates the evolution of purity during time propagation. We observe a decrease in $P [\rhoe (t)]$ and $P [\rhoph (t)]$ over time, which we attribute to the local dephasing of the photoelectron wavepacket as it scatters off the perturbers. Notably, following the ramp-up of the laser envelope, the purity decay is well-characterized by an exponential model
\begin{equation}
\label{eq:purity_fit}
    P [\rho ( t) ] \sim \gamma \left[ \rme^{- (t-t_0)/t^{\star}} - 1 \right] + 1 .
\end{equation}
Within this framework, the fit parameters provide a clear physical picture of the system's transition to a mixed state.
The characteristic timescale of decoherence is $t^{\star}$, the smaller $t^{\star}$ the faster the dissipation of quantum coherences.  The purity floor is $\gamma$, as $P [\rho (\infty)] = 1-\gamma$. Thus, $\gamma = 1$ represents maximal decoherence where the photoelectron state evolves into a purely classical statistical mixture, while $\gamma < 1$ represents residual coherence that can be maintained by the portion of the wavefunction remaining in the bound state.

A numerical fit to the data provides the characteristic parameters for the purity of $P [\rhoe (t)]$ and $P [\rhoph (t) ]$. For the total density matrix $P[\rhoe (t)]$ (solid green curve, Fig.~\ref{fig:purity_time}), we find $\gamma = 0.63$, $t^{\star} = 7.43$ fs and $t_0 = 7.53$ fs. In contrast, the photoelectron subsystem $P [ \rhoph (t)]$ (dotted green curve) yields $t_0 = 6.28$ fs, $t^{\star} = 2.65$ fs and $\gamma = 0.98$, the latter indicate a nearly absolute decoherence rapidly reached.
The purity of the photoelectron subsystem decays more than twice as fast as that of the total wavefunction. This disparity highlights the stabilizing effect of the bound state population. 
Ultimately, the evolution of $P [\rhoe (t)]$ is a competition between free-electron dispersion and retention in the bound-state. The total purity is intrinsically linked to the ionization rate, as the relative weighting between the coherent bound state and the decoherent continuum determines the system's global purity floor.

\begin{figure}
    \centering
    \includegraphics[width=.45\textwidth]{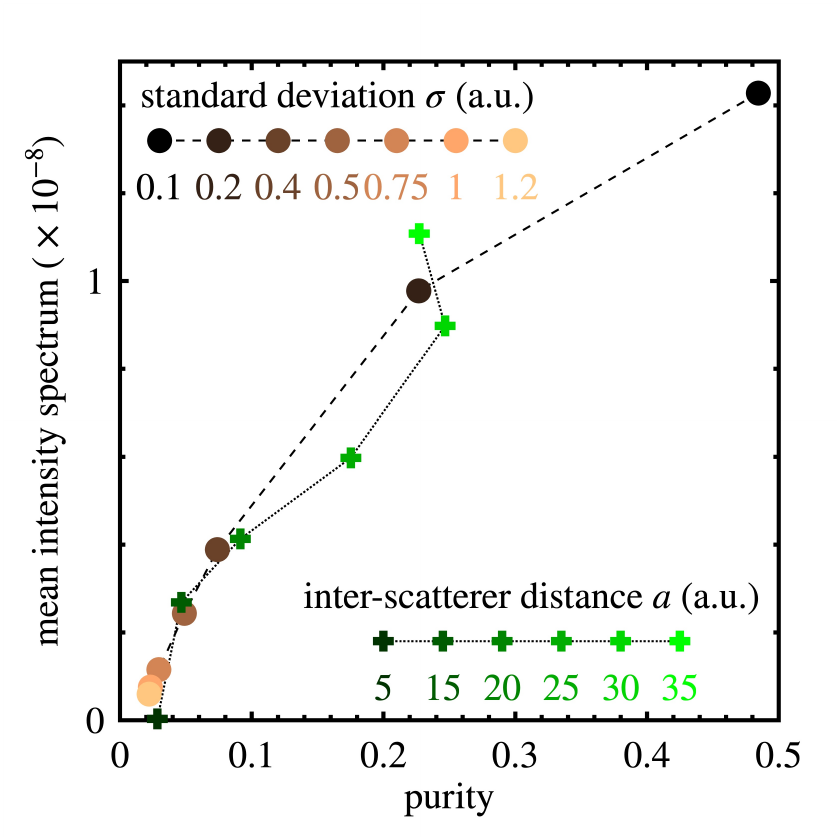}
    \caption{Mean intensity spectrum as a function of the purity of the photoelectron $P [ \rhoph (t)]$ at $t=3.75 \,\Tl$ [see Eq.~\eqref{eq:purity}] of the photoelectron for different values of the standard deviation $\sigma$ (for $a=10$ a.u.) and the inter-scatterer distance $a$ (for $\sigma = 0.1$ a.u.) [see Eq.~\eqref{eq:fdx}]. 
    Mean peak amplitudes are obtained by averaging peak intensities from the 21st to the 227th harmonics (corresponding to the plateau region).}
    \label{fig:sigma_a}
\end{figure}

We subsequently examined the influence of the model's structural parameters in our model, specifically the standard deviation $\sigma$ of the scatterer distribution and the mean inter-scatterer distance $a$ [Eq.~\eqref{eq:fdx}] on HHG. Figure~\ref{fig:sigma_a} displays the mean harmonic peak amplitude (averaged over harmonics 21 to 227) as a function of these parameters. Adjusting these variables effectively controls the structure of the environment and therefore the degree of perturbation of the electron's continuum dynamics, which are central to step (ii) of the three-step model. Increasing $\sigma$ introduces greater spatial disorder by flattening the pair-correlation function displayed in Fig.~\ref{fig:gt}, leading to more stochastically defined scatterer positions. In contrast, increasing the mean distance between perturbers $a$ reduces the frequency of collisions between the photoelectron and the environment, thereby minimizing the disruption of the coherent emission process.

Our results demonstrate that harmonic intensity and state purity $P [\rhoph (t)]$ evolve in tandem with the structural parameters of the environment, aligned with the increase of structural disorder. As the latter increases, the electron’s ensemble-averaged state becomes increasingly mixed, directly suppressing harmonic emission. This transition to a mixed state is fundamentally driven by the cumulative effect of collisions with the scattering centers. Within the context of this 1D study, the mean inter-scatterer distance $a$ serves as a direct proxy for the electron's mean free path, setting the characteristic length scale over which quantum coherence is maintained.

\subsection{Environment-induced decoherence and classical beahvior}

\begin{figure}
    \centering
    \includegraphics[width=.5\textwidth]{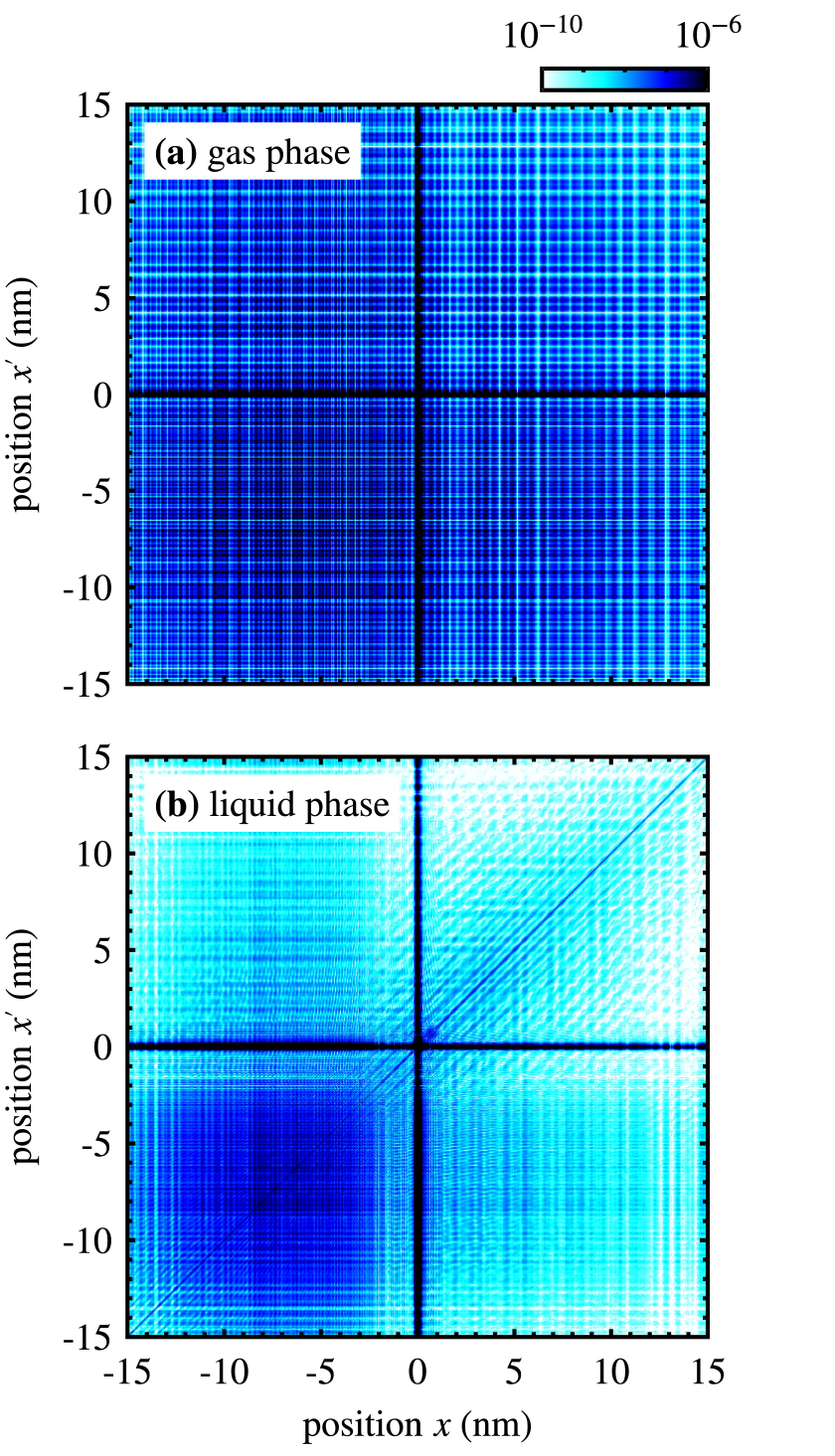}
    \caption{Density matrix $|\rho(x^{\prime},x,t)|^2$ at $3.75 \, \Tl$ after the start of the simulation, for (a) the gas phase and (b) the liquid phase. On each panel, the lines at $x=0$ and $x^{\prime}=0$ indicate the position of the electron in the ground state.}
    \label{fig:rhot}
\end{figure}

The reduction in purity, a direct signature of decoherence, can be explicitly visualized through the spatial representation of the density matrix of the photoelectron
\begin{equation}
    \rho(x,x^{\prime},t) = \bra{x} \hat{\rho} (t) \ket{x^{\prime}} .
\end{equation}
As we have shown, environmental disorder transforms the state of the photoelectron into a statistical mixture. In this context, the degradation of purity is manifested as the suppression of non-diagonal elements ($x \neq x^{\prime}$) in the density matrix. Although decoherence typically arises in open quantum systems via entanglement with an external reservoir, our simulations achieve a similar effect within a closed-system framework. Here, decoherence emerges from the destructive interference of quantum phases across the ensemble of disordered scattering potentials.
To visualize this process, Fig.~\ref{fig:rhot} presents the squared modulus of the density matrix at $t = 3.75\, \Tl$. This snapshot is representative of the system's state across different periods of the laser field. The off-diagonal coherences present in the gas phase vanish in the liquid phase, leaving mainly the diagonal populations and the self-coherences of the ground state (the latter being largely localized and unaffected by the environment). The observed asymmetry in the density matrix reflects the spatial bias imposed by the driving electric field on the tunnel-ionized electron wavepacket during its propagation and subsequent emission.

\begin{figure}
    \centering
    \includegraphics[width=.45\textwidth]{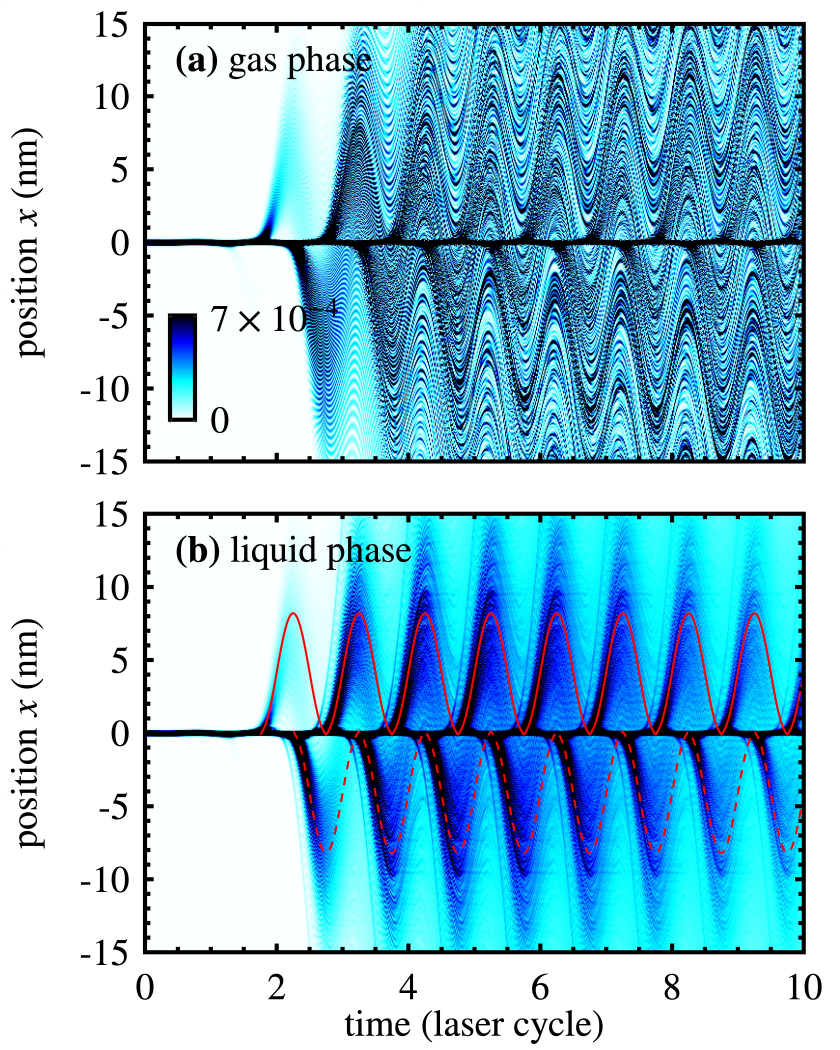}
    \caption{Probability density of presence as a function of time for (a) the gas phase, (b) the liquid phase. In (b), the solid and dashed red lines correspond to periodic orbits of the classical analog system associated with Hamiltonian~\eqref{eq:H0}.}
    \label{fig:density}
\end{figure}

This progressive loss of coherence naturally motivates a deeper investigation of the underlying dynamics of the photoelectron. 
Figure~\ref{fig:density} illustrates this evolution through the spatial probability density $\rho (x,x,t)$, which corresponds to the diagonal elements of the density matrix.
In the gas phase (Fig.~\ref{fig:density}a), the probability density exhibits a complex, delocalized structure. Once the laser envelope completes its ramp-up, various pathways can be identified. However, in the liquid phase, the electronic density becomes remarkably localized along specific pathways, represented by the high-density regions in Fig.~\ref{fig:density}b. Overlaid on this density are the red solid and dashed curves, representing two classical trajectories derived from the Hamiltonian in Eq.~\eqref{eq:H0}. It is evident that the probability density preferentially concentrates around these classical structures.

These trajectories represent specific solutions of Eq.~\eqref{eq:H0} under a constant laser envelope, corresponding to periodic orbits~\cite{Kamor2014, Berman2015, Dubois2022, Floriani2024}. By definition, a periodic orbit returns to its initial phase-space coordinates $\mathbf{z}^{\star}$ after each laser cycle $\Tl$, where $\mathbf{z} = (x,p)$ denotes the phase-space variables. This periodicity is formally expressed by the fixed-point condition
\begin{equation}
\label{eq:condition_PO}
    \boldsymbol{\varphi}_{t_0}^{t_0 + \Tl} (\mathbf{z}^{\star}) = \mathbf{z}^{\star} ,
\end{equation}
where $\boldsymbol{\varphi}_{t_0}^{t}$ denotes the Hamiltonian flow originating from $\mathbf{z}^{\star}$ at time $t_0$.
To locate the specific coordinates $\mathbf{z}^{\star}$ that satisfy~\eqref{eq:condition_PO}, we employ a Newton method, as detailed in~\cite{Dubois2022}. 

The electron trajectories depicted in Fig.~\ref{fig:density} correspond to two distinct periodic orbits related by the internal symmetries of the system. Specifically, the combined centrosymmetry of the potential, $V(x) = - V(x)$, and the half-period symmetry of the laser field, $F(t + \Tl/2) = - F(t)$, ensure that if a trajectory $\boldsymbol{\varphi}_{t_0}^{t} (\mathbf{z}^{\star})$ exists, a symmetric partner also exists. In Fig.~\ref{fig:density}, these two orbits are represented by the spatial component of the flow for $t_0 = 2 \, \Tl$ (solid red line) and $t_0 = 2.5 \, \Tl$ (dashed red line). 
Interestingly, in the current parameter regime, the SFA trajectory~\eqref{eq:xt} under the ionization condition $\cos (\omega \ti) =0$ closely approximates these periodic structures.
A key characteristic of these periodic orbits is their proximity to the origin and their vanishingly small momentum during the peaks of the laser electric field, precisely when tunnel ionization is most probable according to step (i) of the three-step model. Consequently, the newborn photoelectron wave packet enters the continuum in close proximity to these structures in phase space, allowing it to follow their dynamical paths. 
The significance of these periodic orbits in shaping strong-field phenomena has been noted in several recent studies~\cite{Kamor2014, Berman2015, Dubois2022, Floriani2024}. Furthermore, a stability analysis reveals that these paths are hyperbolic, acting as unstable manifolds that drive neighboring trajectories to diverge, thereby framing the localized density as a manifestation of dynamical interference.

The emergence of these unstable classical structures in a perturbed quantum system is strikingly reminiscent of `quantum scars'~\cite{Heller1984, Luukko2016, keski-rahkonen_quantum_2019}, although it is traditionally discussed in the static context of Hamiltonian eigenstates.
Recent work has demonstrated that such localization can be induced by introducing random perturbers into an otherwise integrable system~\cite{Luukko2016, keski-rahkonen_quantum_2019}. However, our system extends this phenomenology to the time-dependent domain, where decoherence plays an active role. In our model, periodic orbits are inherently temporal, and the localization along these classical structures occurs dynamically during the evolution of the wavepacket from the ground state. These observations parallel the landmark findings of Zurek and Habib~\cite{habib_decoherence_1998, zurek_decoherence_1991}, who showed that decoherence, typically induced by a Markovian environment, can trigger a dynamical transition from quantum to classical behavior in chaotic systems.
Critically, in our case, the decoherence is driven by the ensemble averaging over discrete environmental configurations and not by relaxation processes.

Consequently, the strong-field dynamics of an electron in a scattering liquid serves as a compelling example of a quantum-to-classical transition. Here, the transition is mediated by environment-induced decoherence, manifesting itself as the destructive interference of a disordered ensemble within a formally closed quantum system. Unlike traditional open-system models that rely on an external reservoir, our results demonstrate that the intrinsic stochasticity of the liquid phase is sufficient to suppress quantum delocalization, effectively forcing the electron to follow the skeleton of its underlying classical periodic orbits.

\section{Conclusions \label{sec:conclusion}}

In this work, we have investigated the HHG response of an atom within a disordered medium using a one-dimensional stochastic model. Our results demonstrate that the environment fundamentally reshapes the HHG intensity spectrum, most notably through the emergence of a secondary plateau. This feature, which is consistent with recent theoretical and experimental observations~\cite{Mondal2025}, is quantitatively attributed to off-site recombinations on neighboring perturbers. These trajectories allow the electron to reach return kinetic energies that significantly exceed the traditional gas-phase limit of $3.17 \ \Up$.

Central to our findings is the characterization of the photoelectron’s transition from a pure quantum state to a statistical mixture. We have shown that the intrinsic stochasticity of the liquid environment, where each photoelectron encounters a unique spatial configuration, induces robust dynamical decoherence during the propagation step of the three-step model. This decoherence arises from the interplay between various pathways, including off-site recombinations and backscattering events. In the reduced density matrix, this process is manifested by the rapid suppression of off-diagonal coherences and a corresponding power-law decay in system purity.

As a consequence of this decoherence, the electronic probability density undergoes a quantum-to-classical transition, localizing around the unstable, hyperbolic periodic orbits of the classical Hamiltonian. This observation extends the foundational concepts of Heller and Lissajous scars to a regime beyond the constraints of closed-system spectral theory. By operating in a spatially unbounded system without traditional relaxation, we demonstrate that structural disorder alone is sufficient to drive the emergence of \textit{dynamic scars} during the evolution from the ground state. This reveals a profound link between field-driven electron dynamics and the statistical nature of the surrounding medium, potentially providing a new framework for understanding strong-field phenomena in condensed phases.

%\bibliographystyle{apsrev4-1}
%\bibliography{Bibliostage}


\end{document}